\documentclass[twocolumn, aps, prx, floatfix]{revtex4-2}
\pdfoutput=1
\usepackage[T1]{fontenc}
\usepackage{graphicx}
\usepackage[dvipsnames]{xcolor}
\usepackage{hyperref}
\hypersetup{
	colorlinks=true,
	linkcolor=Maroon,
	citecolor=ForestGreen,
	urlcolor=MidnightBlue,
}

\begin{document}


\newcommand{\icsdwebshort}[1]{\href{https://www.topologicalquantumchemistry.fr/\#/detail/#1}{#1}}
\newcommand{\icsdweb}[1]{\href{https://www.topologicalquantumchemistry.fr/\#/detail/#1}{ICSD #1}}
\newcommand{\webNoICSD}{\url{https://www.topologicalquantumchemistry.org/}}
\newcommand{\webTQC}{\href{https://www.topologicalquantumchemistry.fr/}{Topological Quantum Chemistry}}
\newcommand{\webchecktopmat}{\href{https://www.cryst.ehu.es/cryst/checktopologicalmagmat}{Check Topological Mat}}
\newcommand{\identify}{\href{www.cryst.ehu.es/cryst/identify_group}{IDENTIFY GROUP}}
\newcommand{\webflatband}{\href{https://www.topologicalquantumchemistry.fr/flatbands}{Materials Flatband Database}}
\newcommand{\flatwebdirectlink}[2]{\href{https://www.topologicalquantumchemistry.fr/flatbands/index.html?ICSD=#1}{#2}}
\newcommand{\icsdflatweb}[1]{\href{https://www.topologicalquantumchemistry.fr/flatbands/index.html?ICSD=#1}{ICSD #1}}

\newcommand{\bcsidwebshort}[1]{\href{https://www.topologicalquantumchemistry.fr/magnetic/index.html?BCSID=#1}{#1}}
\newcommand{\bcsidweblong}[1]{\href{https://www.topologicalquantumchemistry.fr/magnetic/index.html?BCSID=#1}{BCSID #1}}

\newcommand{\webmaterialsproject}{\href{https://materialsproject.org}{Materials Project}}

\newcommand{\webscnims}{\href{https://mits.nims.go.jp/en}{NIMS Materials Database}}

\newcommand{\bcslong}{\href{https://www.cryst.ehu.es/}{\emph{Bilbao Crystallographic Server}}}

\title{Line-graph-lattice crystal structures of stoichiometric materials}
\author{Christie S. Chiu}
\affiliation{
	Department of Electrical Engineering, Princeton University,
	Princeton, New Jersey, 08544, USA}
\affiliation{
	Princeton Center for Complex Materials, Princeton University,
	Princeton, New Jersey, 08540, USA}
\author{Annette N. Carroll}
\affiliation{
	Department of Physics, Princeton University,
	Princeton, New Jersey, 08544, USA}
\author{Nicolas Regnault}
\affiliation{
	Department of Physics, Princeton University,
	Princeton, New Jersey 08544, USA}
\author{Andrew A. Houck}
\affiliation{
	Department of Electrical Engineering, Princeton University,
	Princeton, New Jersey, 08544, USA}
\date{\today}

\begin{abstract}
The origin of many quantum-material phenomena is intimately related to the presence of flat electronic bands.
In quantum simulation, such bands have been realized through line-graph lattices, a class of lattices known to exhibit flat bands.
Based on that work, we conduct a high-throughput screening for line-graph lattices among the crystalline structures of the \webflatband\ and report on new candidates for line-graph materials and lattice models.
In particular, we find materials with line-graph-lattice structures beyond the two most commonly known examples, the kagome and pyrochlore lattices.
We also identify materials which may exhibit flat topological bands.
Finally, we examine the various line-graph lattices detected and highlight those with gapped flat bands and those most frequently represented among this set of materials.
With the identification of real stoichiometric materials and theoretical lattice geometries, the results of this work may inform future studies of flat-band many-body physics in both condensed matter experiment and theory.
\end{abstract}

\maketitle

\section{Introduction}

Within a dispersionless band of a crystalline solid, electrons have diverging effective mass and localized wavefunctions can remain localized, notably in the absence of disorder.
The inclusion of Coulomb repulsion then gives rise to strongly interacting many-body systems, which have been predicted to exhibit phenomena ranging from ferromagnetism \cite{Mielke1991, Mielke1991a, Tasaki1992, Tasaki2008} and flat-band many-body localization \cite{Danieli2020a, Daumann2020, Kuno2020, Khare2020, Roy2020, Orito2021}, to unconventional superconductivity \cite{Xie2020, Hu2019, Julku2020, Hazra2019} and zero-magnetic-field fractional quantum Hall states \cite{Neupert2011, Wang2012, Regnault2011, Sun2011}.
Experimental work, too, has targeted flat-band physics, for example through quantum simulation on various platforms such as photonics \cite{Leykam2018, Baboux2016, Mukherjee2018, Ma2020a}, quantum circuits \cite{Kollar2019a, Hung2021}, and ultracold atoms \cite{Jo2012, He2020}, as well as on materials such as magic-angle twisted bilayer graphene and twisted bilayer transition metal dichalcogenides \cite{Bistritzer2011,Marchenko2018,Zhang2020,IqbalBaktiUtama2021,Lisi2021}.
	
\begin{figure}[t!]
\centering
\includegraphics{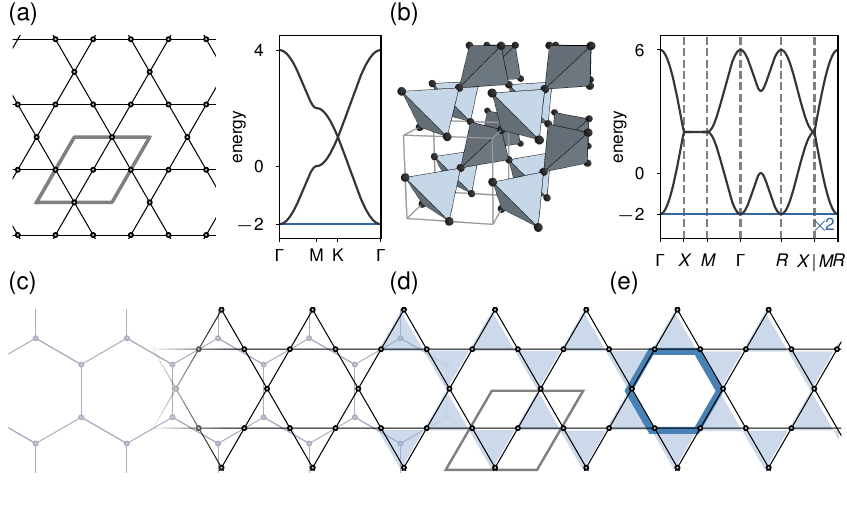}
\caption{\textbf{(a)} The kagome lattice and \textbf{(b)} a pyrochlore-like lattice, and their band energies through high-symmetry points of their respective Brillouin zones. The kagome lattice is the line graph of the honeycomb lattice, and the pyrochlore lattice is the line-graph of the diamond lattice. Under the tight-binding model with $s$-orbital-like hopping of amplitude $1$, these lattices exhibit exactly flat bands at eigenvalue $-2$. Lattice sites are denoted with circles, and hopping between them shown by lines. Unit cells are outlined in gray. For pyrochlore, tetrahedra are colored to aid visibility of the lattice structure. \textbf{(c)} A honeycomb lattice (left) with the kagome overlaid (right), highlighting the line-graph construction connecting the two. \textbf{(d)} The Krausz-$(2, 1)$ partition for the kagome lattice. Here cliques of size $3$, which look like triangles, are highlighted in blue. Because each vertex is part of at most two of these cliques, and each edge is part of exactly one clique, this is indeed a valid Krausz partition. The number and arrangement of cliques per unit cell (outlined in gray) characterizes the lattice, see main text. We also note that the tetrahedra coloring of (b) represents a Krausz partition for the pyrochlore lattice. \textbf{(e)} Line-graph lattices in 2D can be further characterized by considering faces of the lattice, here outlined in blue. See main text for details.}
\label{fig:linelattices}
\end{figure}

Certain families of lattices are known to host flat bands.
For example, bipartite lattices have flat bands at the center of their spectra, with band degeneracy equal to the difference in number of sites per unit cell in each sublattice; the Lieb lattice is a well-known example \cite{Lieb1989}.
Additionally, mathematical generators of flat-band lattices have been proposed \cite{Flach2014, Xu2020}.
Specific lattices have also been identified to host flat bands, including the kagome \cite{Syozi1951} and pyrochlore lattices \cite{Subramanian1983}, see Figure \ref{fig:linelattices}(a).
These are both examples of line-graph lattices, though their flat bands are not gapped \cite{Mielke1991, Mielke1991a}.

Line graphs are graphs (a set of vertices connected by edges) that reflect the adjacency between edges of another graph, which we term the root graph, see Figure \ref{fig:linelattices}(b).
More specifically, every edge in the root graph is represented by a vertex in its line graph, and edges in the line graph connect vertices arising from incident edges in the root graph.
The adjacency matrix of a line graph can be shown to have $-2$ as its lowest eigenvalue \cite{Cvetkovic2004}.
Through the addition of discrete translation invariance, (finite-size) line graphs can be extended to line-graph lattices.
Correspondingly, for dimensions $D>1$ the associated tight-binding Hamiltonian with amplitude-$1$ hopping exhibits one or more exactly flat band or bands at the bottom of the spectrum, with eigenvalue $-2$ \cite{Kollar2019}.

Line-graph lattices have emerged as a means for generating flat bands within the field of quantum simulation with superconducting circuits.
In particular, lattices of microwave cavities have been constructed as a path towards simulating condensed matter systems \cite{Kollar2019a, Carusotto2020}.
In such lattices, each cavity acts as a lattice site for photons.
As a result, a circuit layout with cavities on every edge and capacitive coupling between cavities at each vertex simulates the corresponding line graph.
Stemming from these ideas, the topology of line-graph-lattice flat bands has been examined; line-graph lattices and line-graph lattices with select perturbations have been theoretically shown to have exactly flat or nearly flat fragile topological bands \cite{Chiu2020, Ma2020}.

More generally, the identification and characterization of flat-band lattice models is integral to theoretical and experimental work \cite{Liu2014}.
For example, the kagome lattice is a rich theoretical playground for studying magnetism and resonating valence bond states \cite{Elser1989, Hastings2001}.
More recently, it has inspired materials design to experimentally realize Dirac cones and flat bands \cite{Kang2020a, Kang2020}.
Similarly, the pyrochlore lattice has also been the focus of theoretical simulation and first-principles calculations \cite{Hase2019}.
Much work has been done to identify flat-band materials and classify those with bipartite structure---including split-graph lattices and the Lieb lattice---or with kagome or pyrochlore sublattices \cite{Regnault2021}.
However, prior to this work, it was not known what other crystal structures, if any, are line-graph lattices.

Here we develop and execute a high-throughput screening for crystalline structures that are line-graph lattices.
The materials are from the \webflatband\ \cite{Calugaru2021, Regnault2021}.
This database identifies flat-band materials from the \webTQC\ \cite{Bradlyn2017, Vergniory2021}, which contains most stoichiometric structures from the Inorganic Crystal Structure Database (ICSD).
Of the $55,206$ ICSDs in the Materials Flatband Database, we find $4409$ hosting line-graph lattice crystalline structures, $2970$ of which are not kagome or pyrochlore structures. Our results are publicly available on the \webflatband.
Furthermore, we find over $388$ unique line-graph lattices, verified to be consistent with line graphs by computing the tight-binding model band spectra.
Because line-graph lattices exhibit flat bands due to geometric frustration rather than fine-tuned parameters, these materials and their underlying lattices are particularly promising for materials engineering and design, first-principles theoretical study, and quantum simulation.

\section{Method}
Our algorithm to determine whether a given lattice is a line graph relies upon one key insight: line-graph lattices are composed of fully connected subgraphs, where each bond is part of exactly one subgraph and each site can be a part of at most two subgraphs.
Within graph theory, these fully connected subgraphs are known as cliques.
Such a clique partitioning is called a Krausz-$(2, 1)$ partition \cite{Krausz1943}, which we will refer to as a Krausz partition for simplicity.
If a Krausz partition exists, the resulting graph is a line graph; otherwise, it is not.
The partitions for the kagome and pyrochlore lattices are shown in Figure \ref{fig:linelattices}(d) and (b), respectively.
We note that this is a purely geometric method of identifying line-graph lattices, based solely on the connectivity of sites.
It does not depend on the space symmetry group of the material or occupation of particular high-symmetry points in the lattice (maximal Wyckoff positions).

With this particular consideration in mind, our search proceeds over all Material Flatband Database ICSD entries as follows, see Figure \ref{fig:appxalg}.
First, we determine the lattice structure, given by the connectivity of atomic sites and its dimension.
Following \cite{Regnault2021}, we assume that the hopping between any two atoms depends on their spatial separation and place a cutoff for long bond lengths.
The search is iterated on various cutoff parameters, detailed in Appendix \ref{appx:algorithm}.
We search over the resulting 3D structures to extract lattice geometries with flat bands over the entire three-dimensional (3D) Brillouin zone.
In addition, we search over two-dimensional (2D) structures on the various Miller planes to identify lattices with flat bands along a 2D plane of the Brillouin zone.

Second, we determine whether each structure is a line-graph lattice.
This begins by checking the effective dimensionality of the structure, to analyze only those which are 2D, quasi-2D, or 3D.
Next, we check whether the number of edges (bonds) is below the upper limit for a line-graph lattice, given its number of vertices (sites).
At this point, note that any algorithm to search for a Krausz partition is likely better-suited for finite-sized graphs.
To reduce such a graph without affecting whether it is a line graph, we isolate all of the edges of a single unit cell and their adjacent vertices, such that this graph can be translated by the lattice vectors to construct the entire lattice.
Crucially, while no two edges of this reduced graph are translationally invariant, this is not the case for the vertices.
Upon rearranging these edges and vertices under only lattice vector translations, we create finite-sized graphs which will be line graphs if and only if the original lattice is a line-graph lattice.
In the interest of computational efficiency, at this point we ignore lattices that are too complex, however we estimate the effects of this to be small, see Appendix \ref{appx:algorithm} for details.
The cliques can then be extracted via the Bron-Kerbosh algorithm \cite{Bron1973}, from which the presence or absence of a Krausz partition can be determined.
We can test the success of our algorithm by calculating the tight-binding spectra of our detected line-graph lattices and confirming the presence of exactly flat bands at $-2$ across their respective 3D or 2D Brillouin zones.
Additional details of our algorithm can be found in Appendix \ref{appx:algorithm}.

Finally, we filter the line-graph lattices themselves.
This characterization allows us to identify the prevalence of the kagome and pyrochlore lattices among our extracted materials.
It also allows us to identify other common line-graph lattices that may be of interest for theoretical study.
The most coarse-grained criteria is the dimensionality of the lattice.
Next, we tabulate the sizes of the clique(s) adjacent to each vertex in the unit cell and count the frequency of each clique-size singlet or pair.
Lattices which only differ by integer multiples of these frequencies are grouped together.
This accounts for lattices whose unit cells are different sizes but otherwise equivalent, for example lattices whose unit cells are comprised of two copies of the unit cell of another lattice.

As examples, the characterizations for the pyrochlore and kagome lattices can be seen from Figure \ref{fig:linelattices}(b) and (d).
The depicted pyrochlore unit cell consists of two size-$4$ cliques, with light and dark coloring.
There are four vertices per unit cell, and each is shared by two size-$4$ cliques.
The kagome characterization is similarly simple: the unit cell consists of two size-$3$ cliques, and each of the three vertices per unit cell is shared by two size-$4$ cliques.

In two dimensions, additional characterizations are possible.
In particular, if edge crossings within a clique are permitted (but not across multiple cliques), then the graph can be embedded on a torus.
The concept of ``faces'' of this graph, neglecting the regions bounded by cliques, is then well-defined: they are regions bounded by edges and vertices that contain no other edges or vertices.
For the kagome lattice, as seen in Figure \ref{fig:linelattices}(e), these faces correspond to the non-shaded (non-clique) regions.
They are all hexagons, bounded by six edges and six vertices, as outlined in blue.

As a result, we can determine the size and number of faces per unit cell, the ordered list of clique sizes adjacent to each face, and the two face sizes adjacent to each vertex.
The kagome lattice contains one hexagon (size-$6$ face) per unit cell with six size-$3$ cliques adjacent to it, and two size-$6$ faces are adjacent to each vertex.
As before, lattices which only differ by integer multiples of these frequencies are grouped together.
These attributes fully define the graph, such that the graphs of each group are isomorphic to one another.
By contrast, these characterizations are not possible in 3D.
Our groups of quasi-2D and 3D line-graph lattices may then in fact consist of multiple similar but non-isomorphic lattices, so our cited number of unique line-graph lattices is a lower bound.

\section{Results}

\begin{table}[tb]
	\begin{centering}
		\begin{tabular}{lcccc}
			\hline
			& 2D & quasi-2D & 3D & total\\
			\hline
			unique materials & \hspace{-1.7pt}\begin{tabular}{c}3761\\(6.81\%)\end{tabular} & \hspace{-1.7pt}\begin{tabular}{c}131\\(0.24\%)\end{tabular} & \hspace{-1.7pt}\begin{tabular}{c} 729\\(1.32\%)\end{tabular} & \hspace{-1.7pt}\begin{tabular}{c} 4409\\(7.99\%)\end{tabular}\\
			\hspace{-1.7pt}\begin{tabular}{l} not kagome or \\ pyrochlore-like\end{tabular} & 2655& 129 & $\geq$340 & $\geq$3053\\
			\hspace{-1.7pt}\begin{tabular}{l} gapped \\ (tight-binding model)\end{tabular} & 273 & 7 & 120 & 398\\
			$S$-matrix compatible & 5 & 42 & 504 & 551\\
			\hline
		\end{tabular}
		\caption{Of the $55,\!206$ ICSD entries of the Materials Flatband Database, here we tabulate the number of unique entries exhibiting lattice structures that are line-graph lattices. Percentages are taken relative to the entire set of Materials Flatband Database entries. We analyze structures either by taking a cut through a Miller plane (2D) or by keeping the entire 3D structure; quasi-2D ICSDs arise from 3D structures without tunneling along one spatial direction. Of these ICSDs, we also note the number with lattices that are not the kagome or pyrochlore lattices; have gapped flat bands; or are conducive to the $S$-matrix method (see main text for additional details). Some ICSDs are represented in multiple columns and some give rise to multiple line-graph-lattice structures that differ in the above characteristics.}\label{table:summary}
	\end{centering}
\end{table}

One may not expect to find many crystal structures that are line-graph lattices.
As these lattices are fully comprised of cliques, they contain clusters of atomic sites with all-to-all tunneling of equal amplitude---a feature which seems relatively uncommon.
Indeed, under criteria identifying different features of the kagome and pyrochlore lattices from those examined here, related work has identified just over $11\%$ and $3\%$ of Flatband Materials Database entries hosting kagome and pyrochlore sublattices, respectively \cite{Regnault2021} .

\begin{figure}[t!]
	\centering
	\includegraphics{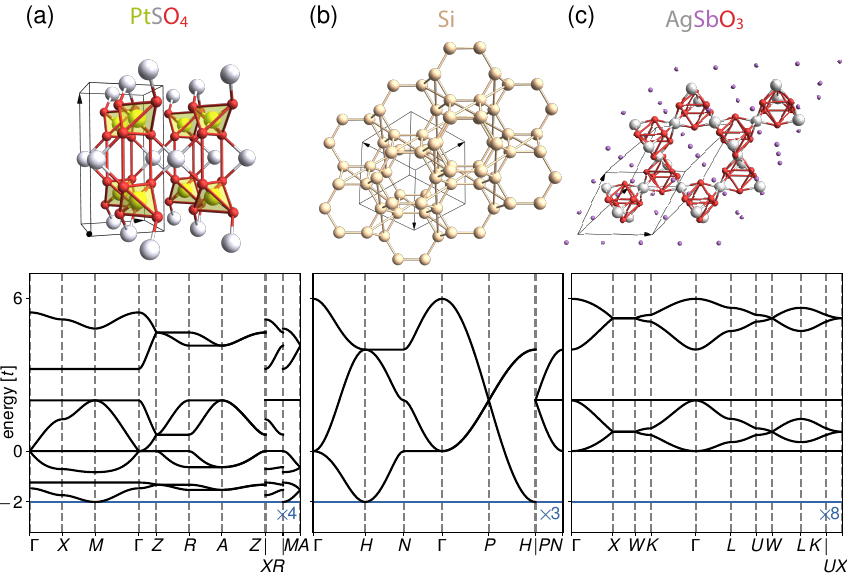}
	\caption{Crystal and band structures of select 3D line-graph-lattice ICSDs, \textbf{(a)} PtSO$_4$ (\icsdflatweb{671491}), \textbf{(b)} Si (\icsdflatweb{189392}), and \textbf{(c)} AgSbO$_3$ (\icsdflatweb{25541}). In (a), the clique partition is shown via the colored tetrahedra, plus the size-$3$ (triangle) cliques between two oxygen and one sulfur atom. Because there is an additional atom in the center of each tetrahedron, those cliques are of size $5$. The partition for (b) consists entirely of size-$4$ cliques. The partition for (c) also consists of cliques of size $4$, but they are arranged differently and each consists of three oxygen atoms and one silver atom. The antimony atoms are each cliques of size $1$, as there are no bonds to other atoms. In all subfigures, unit cells are outlined in black and flat bands in the spectra are highlighted in blue, with the flat-band degeneracy noted.}
	\label{fig:exs3D}
\end{figure}

The summary of our results is in Table \ref{table:summary} and our identified line-graph materials and lattices can be found in the \webflatband.
Among the $55,206$ ICSD entries screened, we find a select set of unique ICSDs with line-graph crystal structures.
Of these, the line graphs are 3D in $729$ ICSDs, quasi-2D in $131$ ICSDs, and lie on a 2D Miller plane in $3761$ ICSDs.
Among 3D lattices, $443$ ICSDs are pyrochlore-like.
Here, this means that the lattice structure is comprised entirely of size-$4$ and size-$5$ cliques where the cliques of size $4$ ($5$) correspond to (center-occupied) tetrahedra, each with all-to-all hopping between the $4$ ($5$) sites.
Their flat bands are also ungapped.
We note that this is an upper bound on the ICSDs which have a pyrochlore lattice structure, as there may exist lattices that fit the above criteria but are not isomorphic to the pyrochlore lattice, for example the one in Figure \ref{fig:exs3D}(b).
There may also be ICSDs that, for different bond cutoffs, create distinct line-graph lattices.
Within these, both pyrochlore-like and non-pyrochlore-like lattices may be represented.
This subtlety also extends to the other characteristics we consider.
Regarding lower dimensions, $2$ ICSDs have the kagome lattice in their quasi-2D layered structure and $1329$ ICSDs have the kagome lattice on at least one of their Miller planes.
Generally speaking, the pyrochlore and kagome lattices, and those of similar clique compositions, are highly represented among the line-graph lattice structures.
These results reflect the fact that these two particular lattices are well-known within the condensed matter community.

\begin{figure}[t]
\centering
\includegraphics{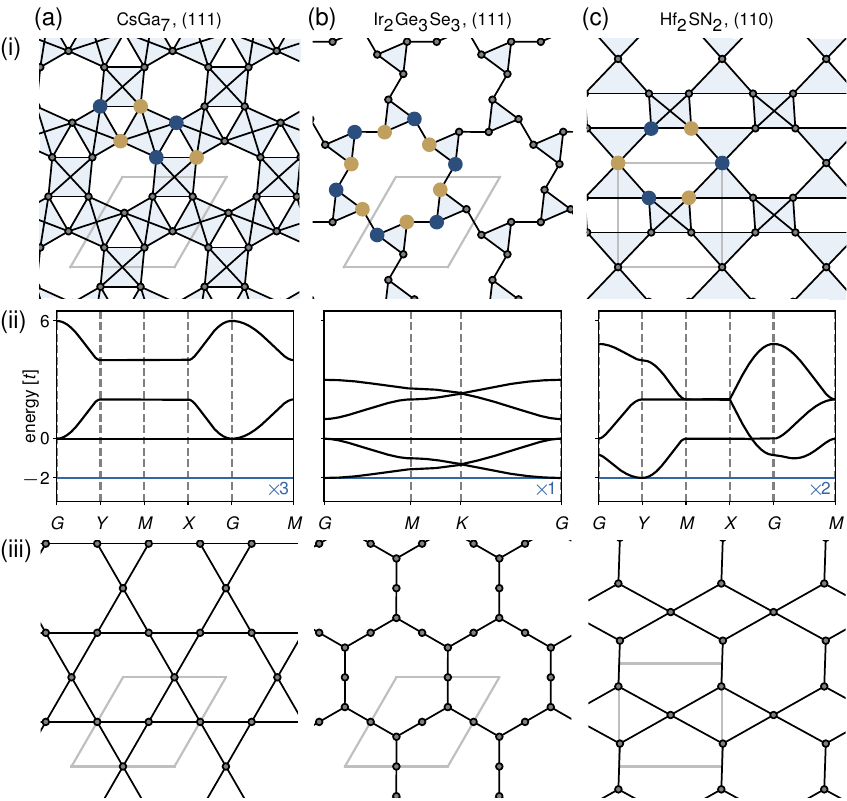}
\caption{\textbf{(i)} Crystal structures and representative compact localized state, \textbf{(ii)}  band structures, and \textbf{(iii)} root graphs of 2D line-graph lattices coming from \textbf{(a)} CsGa$_7$ (\icsdflatweb{102864}) along Miller plane $(1 1 1)$, \textbf{(b)} Ir$_2$Ge$_3$Se$_3$ (\icsdflatweb{636733}) along $(1 1 1)$, and \textbf{(c)} Hf$_2$SN$_2$ (\icsdflatweb{250915}) along $(1 1 0)$. (a) is the line graph of the kagome lattice, (b) is the line graph of the split graph of the honeycomb lattice, and (c) is the line graph of a tiling of hexagons and squares. In (i), unit cells are outlined in grey and the cliques of the clique partition are shaded in light blue. The compact localized state is indicated with real-valued amplitudes on the colored sites, where navy (gold) sites indicate positive (negative) amplitude and all amplitudes are equal in magnitude. In (ii), flat bands are highlighted in blue, with flat-band degeneracy noted.}
\label{fig:exs2D}
\end{figure}

For the majority of these lattices, the flat bands at $-2$ are not gapped; however, $120$ 3D, $7$ quasi-2D, and $273$ 2D ICSDs do exhibit gapped flat bands, with gaps up to $2$ in units of the tunneling amplitude.
In Figures \ref{fig:exs3D} and \ref{fig:exs2D} we highlight a few examples with and without gapped bands, showing their crystal structure and tight-binding spectra along high-symmetry lines.
Figures \ref{fig:exs3D}(c) and \ref{fig:exs2D}(a) provide examples of 3D and 2D lattices, respectively, which have the maximal gap size found.
We additionally include the Krausz partitions and root graphs for the 2D lattices in Figure \ref{fig:exs2D}.

The gappedness and degeneracy of these flat bands can be understood by counting the number of linearly independent flat-band eigenstates, termed ``compact localized states'' \cite{Sutherland1986, Aoki1996}.
Within the subspace of gapped flat bands, the number of linearly independent compact localized states per unit cell equals the flat-band band degeneracy.
If the band is instead ungapped, there will be additional eigenstates at the flat-band energy, each indicating a band touching from dispersive bands \cite{Bergman2008, Kollar2019, Chiu2020}.
Figure \ref{fig:exs2D}(a) shows representative flat-band eigenstates for our examples.
We find lattice structures with band degeneracies from $1$ up to $24$.
Generally speaking, lattices with smaller band degeneracies also have fewer sites per unit cell and therefore may be more amenable to theoretical study.

\begin{figure}[t!]
	\centering
	\includegraphics{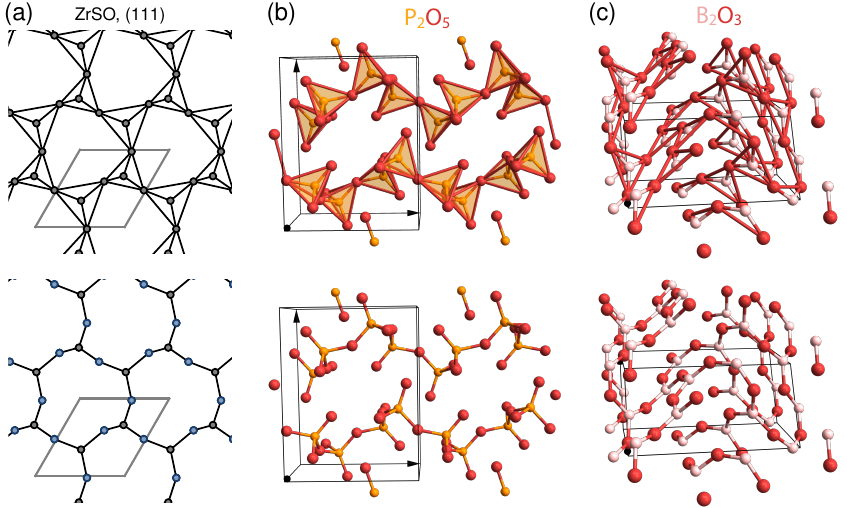}
	\caption{Example ICSDs to which the $S$-matrix method applies: \textbf{(a)} ZrSO (\icsdflatweb{31721}) along Miller plane $(1, 1, 1)$, \textbf{(b)} P$_2$O$_3$ (\icsdflatweb{36066}), and \textbf{(c)} B$_2$O$_3$ (\icsdflatweb{79698}). The upper row indicates the lattice structure, which is a line graph, while the lower row shows how the structure can be decomposed into a bipartite lattice after omitting a subset of bonds.}
	\label{fig:Smatrix}
\end{figure}

Given that our tight-binding model na\"ively assumes $s$-orbital tunneling and no spin-orbit coupling, we next identify the set of line-graph lattices that can be analyzed using the $S$-matrix method \cite{Calugaru2021, Regnault2021}.
If a lattice can be decomposed into a bipartite lattice of sublattices $A$ and $B$, where $A$ contains a greater number of sites than $B$, then the $S$-matrix method applies.\footnote{While some lattices like the kagome are not strictly speaking bipartite, they can be obtained as a limit case of the $S$-matrix method \cite{Regnault2021}}
Then, if the $A$ sublattice is only weakly perturbed by the $B$ sublattice orbitals, the bipartite lattice can be expected to exhibit flat topological bands, regardless of its orbital composition and presence or absence of spin-orbit coupling \cite{Calugaru2021}.
As shown in the examples of Figure \ref{fig:Smatrix}, we find that any given line-graph lattice has a bipartite decomposition if in its Krausz partition, at least one vertex per clique does not belong to any other cliques.
The $B$ sublattice is given by those vertices belonging to only one clique, while the $A$ sublattice is given by the remaining vertices.
Upon omitting a subset of bonds in the original line-graph lattice, as shown in the lower row of Figure \ref{fig:Smatrix}, the lattice can be made bipartite.
Incidentally, this subset consists of the longest bonds in the lattice, implying an shortened effective bond-length cutoff.
In total, we find $504$ ICSDs in 3D, $42$ in quasi-2D, and $5$ in 2D that are amenable to a bipartite decomposition and therefore may be analyzed using the $S$-matrix method.

\begin{figure}[htp]
	\centering
	\includegraphics{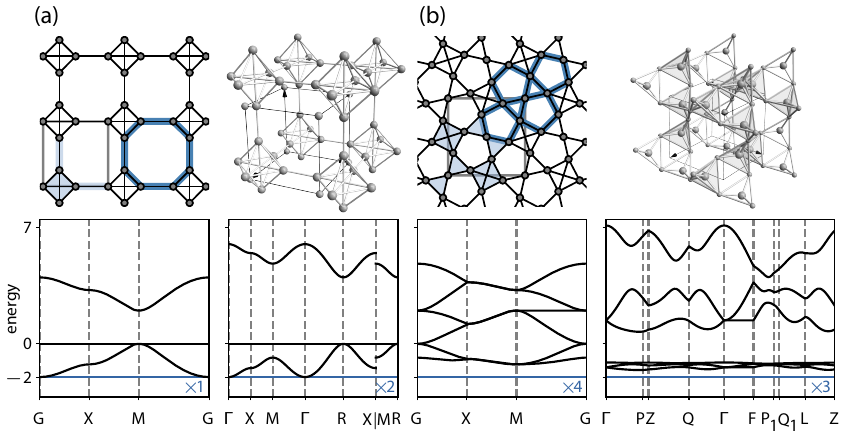}
	\caption{\textbf{(a)} The line graph of the Lieb lattice and of the 3D Lieb-lattice analog. Apart from the kagome and pyrochlore-like lattices, these are two of the most commonly represented lattices in 2D and 3D among the $4409$ line-graph-lattice ICSDs found, seen in $416$ and $71$ unique ICSDs, respectively. \textbf{(b)} The line graph of the Cairo tiling and a non-pyrochlore lattice of center-occupied tetrahedra. Unlike the kagome and pyrochlore lattices, the flat bands of these lattices are gapped. The former is seen in $75$ ICSDs; the latter is seen in $86$ ICSDs. In the top row, unit cells are outlined in grey; in the bottom row, flat bands are highlighted in blue and labeled by their degeneracy. Their characterizations are included in the main text. For the 2D lattices, the cliques and faces referred to in the characterization are highlighted and outlined, respectively.}
	\label{fig:commonlattices}
\end{figure}

Many line-graph-lattice materials have the same lattice structure, which may indicate particular line-graph lattices of interest.
Table \ref{table:lattice} contains our results determining the number of unique line-graph lattices represented in 2D, quasi-2D, and 3D.
Of these lattices, the kagome and pyrochlore-like lattices appear most frequently; over $30\%$ of our unique line-graph materials exhibit one of these structures.
However, we also find a high degree of representation for the lattices shown in Figure \ref{fig:commonlattices}(a).
They are the line graph of the Lieb lattice and of the line graph of the 3D Lieb-lattice analog.

We characterize these three commonly represented line-graph lattices as follows.
The line graph of the Lieb lattice is comprised of one size-$4$ clique and two size-$2$ cliques (per unit cell), highlighted in blue in Figure \ref{fig:commonlattices}(a).
Of its four vertices, all four are adjacent to one size-$4$ clique and one size-$2$ clique.
This lattice also has one octagon (size-$8$) face, outlined in blue, around which there are four size-$4$ and four size-$2$ cliques in alternating fashion, and each vertex is adjacent to two size-$8$ faces.
The line graph of the 3D Lieb-lattice analog (c) has one size-$6$ clique and three size-$3$ cliques per unit cell.
All six of its vertices are adjacent to one size-$6$ and one size-$2$ clique.

Finally, we highlight extracted line-graph lattices which exhibit gapped flat bands, in contrast to the ungapped flat bands of the kagome and pyrochlore lattices.
In Figure \ref{fig:commonlattices}(b), we present the line graph of the Cairo tiling and a non-pyrochlore lattice of center-occupied tetrahedra.
Four size-$3$ cliques and two size-$4$ cliques make up a unit cell of the line graph of the Cairo tiling, where two vertices are adjacent to two size-$3$ cliques and eight are adjacent to one of each size.
There are four pentagon (size-$5$) faces, outlined in blue, around which the size-$3$ and $4$ cliques are interspersed.
Each vertex is adjacent to two size-$5$ faces.
Interestingly, the non-pyrochlore lattice is very similar to pyrochlore in that all of its attributes under our filtering algorithm are identical.
Yet, its center-occupied tetrahedra (size-$5$) cliques are arranged in such a way that the tight-binding band spectrum exhibits gapped flat bands.
This lattice exemplifies how specific lattice geometries may lead to qualitatively different behavior, even among materials which are stoichiometrically similar.

\begin{table}[tb]
	\begin{centering}
		\begin{tabular}{l@{\hspace{12pt}}c@{\hspace{10pt}}c@{\hspace{10pt}}c@{\hspace{10pt}}c}
			\hline
			& 2D & quasi-2D & 3D & total\\
			\hline
			unique lattices & 293& $\geq$60 & $\geq$55 & $\geq$385\\
			gapped & 54 & $\geq$7 & $\geq$20 & $\geq$81\\
			$S$-matrix compatible & 4 & $\geq$9 & $\geq$7 & $\geq$18\\
			\hline
		\end{tabular}
		\caption{Because many ICSDs exhibit the same line-graph-lattice structures, here we tabulate the number of unique line-graph lattices found and further categorize them into the ones which are gapped or the ones to which the $S$-matrix method applies.
		}\label{table:lattice}
	\end{centering}
\end{table}

\section{Discussion}

One of the goals of quantum simulation is to solve quantum mechanical problems which cannot be solved with current classical computation.
Many such open questions exist within condensed matter, and to this end quantum simulation has made great progress on a multitude of experimental platforms.
More specifically, they have provided a mechanism to benchmark and test numerical techniques and theories, giving rise to new intuition and understanding.

Here, we have taken intuition fostered through the development of superconducting-circuit-based quantum simulation and apply it to a search for real-material candidates.
Of the $55,206$ ICSDs examined, almost $8\%$ are found to host line-graph lattices.
A full description per ICSD entry of these line-graph lattices is provided in the \webflatband. These candidates can be probed through condensed matter experiment and may be a starting point for identifying materials that host strongly interacting electrons in flat bands.
This work demonstrates how insights gained from working with synthetic matter can lead to actionable results in the search for new quantum materials. 

Furthermore, from these ICSDs we have found numerous unique line-graph lattices, which give rise to flat bands due to geometric frustration, rather than a fine-tuning of parameters.
Notably, while the kagome and pyrochlore lattices are well-known and prevalent examples, they both exhibit ungapped flat bands in their tight-binding spectra.
We identify additional line-graph lattices and quantify their prevalence.
Of these, we find the line-graph lattices that host gapped flat bands.

Immediate extensions include the development of related algorithms to search for other lattices and families of lattices known to host flat bands.
This includes Tasaki's lattices \cite{Tasaki1998}, lattices that are constructed entirely from cliques but are not line graphs \cite{Tanaka2020}, and decorated or superlattices built from 1D chains \cite{Mizoguchi2019, Lee2020}.
Our search can also be run on a database of monolayer materials.

Of course, the line-graph property of a crystalline structure does not directly indicate that the material itself has a flat band, due to orbital and spin degrees of freedom, varied hopping strengths and next-nearest-neighbor hopping, and disorder.
However, the properties of these structures can be compared to those of ``sister materials'', which have similar composition but are arranged in the root graph structure.
Differences may reveal physics unique to the line-graph flat bands.
Indeed, the band spectra of root-graph lattices and their line graphs differ in that only the line graph exhibits flat bands as its lowest bands, but they can otherwise be quite similar.
The role of lattice geometry may also be disentangled from other degrees of freedom through comparing materials which have the same underlying line-graph lattice, but otherwise differ in their symmetry or other aspects.
More broadly, these newly highlighted lattices are of particular importance given their potential in designing real and synthetic flat-band materials for studies of strongly correlated many-body physics.

\begin{acknowledgments}
We would like to thank B. Andrei Bernevig, Jens Koch, Aavishkar Patel, and Liujun Zou for helpful discussions. The 3D lattice visualizations in Figures \ref{fig:exs3D}, \ref{fig:Smatrix}, and \ref{fig:commonlattices} were created using the Crystallica package, developed by Bianca Eifert and Christian Heiliger of Justus Liebig University Giessen and distributed under the MIT License. We acknowledge support from the Princeton Center for Complex Materials NSF DMR-1420541 and from the ARO MURI W911NF-15-1-0397. N.R. has received funding from the European Research Council (ERC) under the European Union's Horizon 2020 research and innovation programme (grant agreement no. 101020833).
\end{acknowledgments}

\bibliography{linematsearch}

\clearpage
\onecolumngrid
\setcounter{figure}{0}
\renewcommand{\thefigure}{A\arabic{figure}}
\newpage
\appendix

\section{Algorithm}\label{appx:algorithm}

Here we provide additional details on the algorithm used to determine which materials have line-graph-lattice crystalline structures.
Broadly speaking, the algorithm can be split into four parts.
First, we determine the cutoff bond length and optionally take a cut through a Miller plane.
Second, we filter out lattices which have dimensionality below 2D, as well as lattices whose numbers of edges and vertices prohibit a Krausz partition.
Third, we reduce the translationally invariant lattice to a finite-sized graph.
Finally, we extract the cliques present as subgraphs of our graph and determine whether a Krausz$-(2, 1)$ partition exists.

We note that this is not the only way to determine whether a given lattice is a line graph.
For example, there are nine forbidden minimal subgraphs; if a subset of vertices of the graph, combined with all edges of the graph connecting those vertices, creates one of these forbidden subgraphs, then the candidate graph cannot be a line graph \cite{Beineke1970}.
However, our chosen algorithm for line-graph testing offers ways to subsequently characterize the detected line graphs.
It can also be straightfowardly extended and applied to lattices.

\subsection{Parameters}
Following \cite{Regnault2021}, we determine our bond cutoff length through two parameters: a maximum allowed bond length $m$ and an overall multiplicative coefficient $c$.
After calculating the minimum distance $d$ between any two atoms in the lattice, we create bonds between any two atoms which are closer than the distance $c\cdot\mathrm{max}(m, d)$.
We examine all lattices resulting from $m \in \{1.5, 1.8, 2.1, 2.4, 2.7\}$ and $c \in \{1.2, 1.5, 1.7\}$.

For each lattice, we test the full 3D lattice as well as those along each of its Miller planes.
In these planes, bonds between atoms are inherited from the 3D lattice.
Symmetrically redundant Miller planes, identified via the pymatgen package \cite{Setyawan2010, Ong2013, Togo2018}, are omitted.

\subsection{Filtering}
Our first filtering stage consists of examining the connectivities of the vertices to determine whether atomic bonding occurs across two or more dimensions.
For example, if the bond distance is sufficiently small, there may be clusters of atoms which are locally connected, but not connected across unit cells; these lattices are omitted from our search.
Lattices on Miller planes must extend over the full two dimensions, while for full 3D lattices we keep those extending across two or three dimensions.

Regarding the second filtering stage, note that given a lattice with $v$ vertices per unit cell, the minimum number of edges $e$ allowed for a Krausz partition is $e_\mathrm{min} = 0$.
This is the trivial case where each vertex is in its own clique of size $1$.
The maximum number of edges is $e_\mathrm{max} = v \cdot (2v-1)$.
This follows from the fact that a clique with $v'$ vertices has $\frac{v' \cdot (v' -1)}{2}$ edges.
Each vertex can be part of two cliques, but the two cliques need not be discrete.
Thus, the maximal number of edges results from having only a single clique of $v'=2v$ vertices, where each vertex of the unit cell is represented exactly twice.
This single clique then has $v \cdot (2v -1)$ edges.
By applying this very conservative filter, we eliminate lattices with an impractical number of edges.

\subsection{Reduction to finite-sized graph}
\begin{figure}
	\begin{minipage}[c]{0.5\textwidth}
		\includegraphics[width=\textwidth]{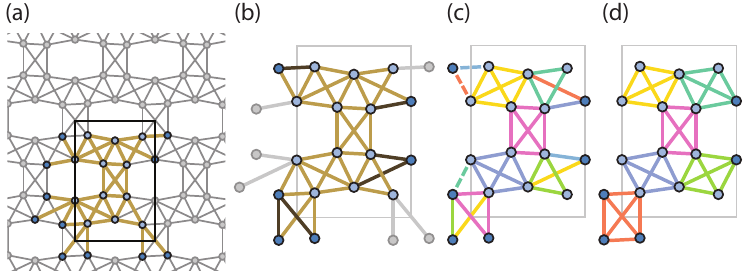}
	\end{minipage}\hfill
	\begin{minipage}[c]{0.45\textwidth}
		\caption{Reduction and Kraus partition search for the line graph of the kagome lattice. \textbf{(a)} Translationally invariant lattice in gray, with isolated edges and vertices (colored) for the initial reduction to a finite-sized graph. \textbf{(b)} Translation of edges along lattice vectors to increase the number of triangle subgraphs. The goal is to create a graph which is a line graph if and only if the original lattice is a line-graph lattice. Gray edges are translated to the dark brown ones, upon which gray vertices are removed. \textbf{(c)} Failed attempt at a Krausz partition, indicated by the colored cliques, with dashed edges indicating ambigous edges, see text for details. \textbf{(d)} Successful Krausz partition upon moving the ambiguous edges. In all subfigures, the outlined rectangle indicates the unit cell and lattice vectors.} \label{fig:appxalg}
	\end{minipage}
\end{figure}

The goal of this step of the algorithm is to convert the lattice into a finite-sized graph, such that the graph is a line graph if and only if the lattice is a line-graph lattice.
Thus, it is sufficient to create a finite-sized graph that, when translated along the lattice vectors, reproduces the entire lattice.
The number of edges of the graph is then given by the number of edges in the unit cell; however, the number of vertices may be greater than the number of sites per unit cell.
As an example, see Figure \ref{fig:appxalg}, which shows the reduction process for the line graph of the kagome lattice.
These additional vertices will have counterparts in the graph, separated by integer numbers of the lattice vectors.
As a result, we first take the lattice adjacency list and represent it as if it were a finite-sized graph, where translationally invariant vertices (in neighboring unit cells) are labeled as unique.

In Figure \ref{fig:appxalg}(a), this is done by considering all sites within a single unit cell, colored in light blue, along with one translationally invariant copy of all edges incident on those sites (gold) and their incident vertices.
While the number of edges is equal to the number per unit cell, this is not the case for the total set of vertices.
The dark blue vertices lie outside of the unit cell outlined in black, \emph{i.e.} each is translationally invariant from one of the light blue vertices.
In creating the finite-sized graph, however, these sites are treated as separate vertices.

We then further manipulate the graph by moving edges and vertices by integer multiples of the lattice vectors, to create a graph which is as similar to a line graph as possible.
More specifically, fully connected subgraphs (cliques) consist of many triangles formed by three edges.
Thus our algorithm preferentially translates edges and sets of edges to create clusters of triangles with shared edges.
Figure \ref{fig:appxalg}(b) shows how the original edges of (a) can be moved from the light-grey locations to the dark brown ones.
This reduces the number of vertices with only one incident edge, which is equivalent to reducing the total number of vertices in the graph.

In some cases, even after this optimization there remain ambiguities regarding how a subset of edges should be translated.
In Figure \ref{fig:appxalg}(c), these are shown as the three dashed edges.
This requires that we search for a Krausz partition for each possible combination of translations, exhibiting exponential scaling with the number of ambiguous edges.
As a result, we set an upper limit for the number of possibilities accepted, below which we check for a Krausz partition among all possible arrangements.
Through examining the tight-binding model band spectra, we estimate that this cutoff leaves undetected at most $1$ 2D and $27$ quasi-2D and 3D line-graph materials.

\subsection{Krausz partition search}
Our algorithm determines whether the graph is a line graph by first identifying all maximal cliques of the graph.
This is done with the Bron-Kerbosh algorithm \cite{Bron1973}.
Then, we determine whether a subset of these cliques fulfills the conditions of a Krausz partition: all edges are part of exactly one clique, while each vertex is part of at most two cliques.
As a subtlety, there are cases in which a clique of size $3$ ought to be instead represented as a clique of size $2$, with two edges and a vertex which are part of other cliques; we necessarily take cases such as these into consideration.

Although the graph in Figure \ref{fig:appxalg}(c) does not admit a Krausz partition, as seen from an attempt to color the cliques of the graph.
Note that in (c), colors are reused for visibility and disjoint subgraphs of the same coloring represent separate cliques.
Multiple vertices of this graph are contained within more than two cliques.
In contrast, upon translating the three ambiguous (dashed) edges, in (d) we find that we can create a graph which is indeed a line graph.
Notice that because the number of vertices is still greater than the number per unit cell, some vertices are represented in duplicate.
As a result, in determining whether a Krausz partition exists, the number of adjacent cliques must be summed for these equivalent vertices.

\end{document}